\documentclass[fullpage,11pt]{article}

\usepackage{amsfonts}
\usepackage{bbm}
\usepackage{mathrsfs}
\usepackage{amsmath}
\usepackage{amsthm}
\usepackage{amssymb}
\usepackage{graphicx,float}
\usepackage{comment}

\usepackage[subfigure]{graphfig}

\begin{document}

\newtheorem{coro}{Corollary}
\newtheorem{defi}{Definition}
\newtheorem{exam}{Example}
\newtheorem{lemm}{Lemma}
\newtheorem{algo}{Algorithm}
\newtheorem{prop}{Proposition}
\newtheorem{theo}{Theorem}
\newtheorem{property}{Property}
\begin{titlepage}
\title{\LARGE On the Complexity of Envy-Free Cake Cutting\vspace{0.6cm}}

\author{
    {Xiaotie Deng} \\ {\small Department of Computer Science} \\ {\small City University
    of Hong Kong} \\ {\small Hong Kong SAR, P.R.China} \\ {\small
    deng@cs.cityu.edu.hk}
  \and
  {Qi Qi} \\ {\small Department of Management Science and Engineering } \\ {\small Stanford University} \\ {\small Stanford, California, USA} \\ {\small kaylaqi@stanford.edu}
  \and
  {Amin Saberi\footnote{{\bf Acknowledgements:} Amin Saberi would like to thank Arash
Asadpour as well as the organizers and participants of Dagstuhl's
Fair Division workshop for valuable discussions.}}\\ {\small
Department of Management Science and Engineering }\\ {\small
Stanford University} \\{\small Stanford, California, USA}\\ {\small
saberi@stanford.edu} }
\date{}\vspace{0.2cm} \maketitle
\end{titlepage}

\begin{abstract}
We study the envy-free cake-cutting problem for $d+1$ players with
$d$ cuts, for both the oracle function model and the polynomial time
function model. For the former, we derive a
$\theta(({1\over\epsilon})^{d-1})$ time matching bound for the query
complexity of $d+1$ player cake cutting with Lipschitz utilities for
any $d> 1$. When the utility functions are given by a polynomial
time algorithm, we prove the problem to be PPAD-complete.

For measurable utility functions, we find a fully polynomial-time
algorithm for finding an approximate envy-free allocation of a cake
among three people using two cuts.
\end{abstract}

\section{Introduction}
%

Suppose you have a cake represented by the interval $([0,1])$, and
you would like to divide it among $n$ persons fairly. Each person
may have a different opinion as to which part is more valuable.
There is a big literature on this problem in economics, political
science and computer
science~\cite{Steinhaus1948,Steinhaus1949,evenpaz,BT,k2,vangelis,kk,asadpour,ws}.
In particular, it is proved using a fixed-point argument that this
problem has an envy-free solution~\cite{Stromquist1980,simmons,Su}.
In other words, it is possible to cut a cake into $n$ pieces
($X=\{l_0, l_1,\cdots, l_{n-1}\}$ from left to right along $[0,1]$)
using $n-1$ cuts and to allocate one piece to each person (player
$i$ assigned piece $l_{\pi(i)}$ for permutation $\pi(i)$) so that
everyone values his or her assigned piece no less than any other
piece.
The question is: is there an efficient
algorithm that finds such a cut (called $(n-1)$-cut subsequently) of
the cake?

A related but less demanding solution than an envy-free solution is
that of proportional cuts. That is, each person gets a piece which
he or she values more than $1/n$ of total. Its complexity has
recently completely solved by Edmonds and
Pruhs~\cite{ep1}\cite{ep2}.

For the $d+1$-person envy-free cut problem with exactly $d$ cuts
under our consideration, however, progress in complexity analysis
has been limited. The existence of such a solution was proven by
Stromquist~\cite{Stromquist1980} with a fixed-point argument. His
proof implies that an $\epsilon$-approximation can be found in time
exponential in input size $O(\log{\frac{1}{\epsilon}})$. Let
$N=\frac{1}{\epsilon}$ throughout our discussion.

We establish three main results: 1) When the best choices of the
players are given by polynomial-time algorithms, we prove that the
problem is PPAD-complete. 2) If the choices are given by a
functional oracle, we derive a $\theta(({1\over\epsilon})^{d-1})$
matching bound for the query complexity of cutting a cake for $d+1$
players. Despite a strong connection, mathematical and
complexity-wise, between equilibrium computation and fixed point
computation, this is the first matching query complexity result for
an equilibrium computation problem. We know of no such results for
Nash equilibrium, which is also PPAD-complete with a strong tie with
fixed-point computation. 3) For the special case of measurable
utility functions, we make a simple observation: there is a fully
polynomial-time approximation scheme for finding an approximate
envy-free allocation 2-cut of a cake among three people. We sketch
our approaches as follows.

 \paragraph{PPAD completeness:} First, we capture the
concept of approximation by defining a discrete envy-free cut (set)
$(\pi, X^{(0)}, X^{(1)},\cdots, X^{(d)})$  such that player $i$
prefers the $\pi(i)$-th piece of the d-cut $X^{(j)}$ for some $j$,
and the d-cuts $\{ X^{(0)}, X^{(1)},\cdots, X^{(d)}\}$ are within
$\epsilon$ distance (in $L_{\infty}$ metric) of each other. Such a
solution converges to an exact envy-free cake cut as the distance
bound for the $d$-cuts goes to zero. Using barycentric coordinates,
the $d$-cut can also be represented as $(x_0,x_1,\cdots,x_{d})$ with
$l_0=x_0/N, l_1=x_1/N, \cdots, l_{d}=x_{d}/N$. Note that
$(x_0,x_1,\cdots,x_{d})$ is a point on the standard $d$-dimensional
simplex. To prove that the problem is in PPAD, we reduce it to the
problem of finding a fully colored base cell in a triangulated
Sperner coloring of a $d$-simplex by using Kuhn's
triangulation~\cite{Kuhn}. This can be done by a two-stage process:
labeling and coloring. First, for a $d$-simplex and a Kuhn's
triangulation with vertex set $V$, a labeling $\mathcal{L}:
V\rightarrow \{0,1,\cdots,d\}$ is valid if $\forall X,Y\in V$ and
$X$,$Y$ on the same base cell, $\mathcal{L}(X)\neq \mathcal{L}(Y) $.
Then for any labeled vertex $X$, we define a coloring
$\mathcal{C}:V\rightarrow \{0,1,\cdots,d\}$ such that
$\mathcal{C}(X)=i$ if player $\mathcal{L}(X)$ prefers the $i$-th
piece of the cut $X$. By a mild condition for the utility functions,
$\mathcal{C}$ is a proper Sperner coloring. Therefore, the key point
here is to find a polynomial time labeling rule. We define the
labeling rule as: $\mathcal{L}(X)= \sum_{i=0}^{d}ix_i \mod (d+1)$
with a proof of its validity for Kuhn's triangulation. On the other
hand, we design a reduction based on the 2D BROUWER
problem~\cite{PPAD,CD2005,2DSperner,CD2006} for its proof of
PPAD-hardness.

\paragraph{Matching bound in the Oracle function model:} We derive a
$\theta(({1\over\epsilon})^{d-1})$ time matching bound for the query
complexity of cake cutting  for $d+1$ players with Lipschitz
utilities. The tight upper bound requires a divide-\&-conquer method
that finds a balanced cut of the simplex. It is made possible by
Kuhn's triangulation of the simplex, and our labeling method for the
envy-free cake cutting problem that allows an efficient parity
checking of the boundary. For the lower bound, the results are
obtained by a reduction to the zero-point
problem~\cite{HPV,Tak,ERR,CD2005}. The reduction is achieved in two
steps. First, we reduce the zero-point problem for
direction-preserving functions to the problem of finding a discrete
fixed-point on a hypergrid. In the second step, we prove that the
hypergrid can be embedded into the original $d$-simplex for
cake-cutting such that its coloring can be extended to a proper
Sperner coloring of vertices in the triangulated simplex.

Instrumental to our matching bound for envy-free cake-cutting, we
prove a matching bound for the SPERNER problem for any constant
dimension $d>2$, in the oracle function model. This was an open
problem, while for the case of $d=2$, a tight bound was known by a
lower bound of Crescenzi and Silvestri~\cite{CS} and an upper bound
of Friedl, et al~\cite{FISV}. This matching bound for the SPERNER
problem may have other applications for fixed-point based solutions.

\paragraph{Fully PTAS for three players with measurable utility functions:}
Finally, for the special case of measurable utility functions, we
are able to utilize their monotone properties to construct an
$\epsilon$-approximate envy-free solution in time polynomial in
$\log(\frac{1}{\epsilon})$ for three players.

We still rely on the general approach of branch-\&-bound on parity
but exploit the monotonicity of the best choice along certain
lines of the possible cuts to make an efficient count of the index
along the boundary. Any
player with a measurable utility function would prefer $A$ to $B$
for $B\subseteq A$. Therefore, when one cut is fixed, one of the
three pieces is fixed. Any player's preference on the other two
pieces will change monotonically as another cut changes from left to
right. Using the barycentric coordinate $X=(x_0,x_1,x_2)$, along the
line of fixed $x_0$, let $x_1$ increases from $0$ to $N-x_0$. The
preference of a player's choice will start with the last piece
$l_2$, to $l_0$, and to $l_1$ (one or two of them may be missing).
Similar monotone property holds when $x_2$ is fixed. For each
player, we can find the boundary point along those lines by binary
search and the break point of the choices will can be obtained in
$O(\log N)$ time.

We cut the space along those two directions, so that the choice
function of each individual player will be monotone along those
directions. Because of the monotonicity and using the above
procedure, we  can calculate the indices of edges along those lines
efficiently. Therefore, the indices of the two regions split by the
cut will be decided quickly. We will stay on the region with an odd
index so that we should end with one that is a diamond shape polygon
consisting of at most two base cells. Because of parity, one of two
cells is  a fully colored base triangle. The overall query
complexity and running time will be of  $O(\log^2 N)$.

\section{Triangulation and Index}

Our results are based on Sperner's Lemma and its generalizations
using the concept of the index of a region~\cite{Todd}. These
results have been fundamental in discrete fixed point computation
(see, e.g., ~\cite{Scarf} and ~\cite{Todd}) and establishing
Brouwer's fixed point theorem~\cite{Brouwer}.

Starting at two dimensions, a triangular grid of scale $1$ is
 an ordinary triangle $\Delta$
 which has three vertices and
one base cell. A triangular grid of scale $N$ places
$\mbox{N-1}$ equally spaced line segments parallel to each of the
three edges of $\Delta$ and divides the triangle into $N^2$ base
cells.

We refer to the three vertices of $\Delta$ as {\em corner
vertices}, and denote them by $D_0$, $D_1$, and $D_2$. The vertices
along the edges of $\Delta$ are referred to as {\em boundary
 vertices}. Other vertices are referred to as {\em internal vertices}. Edges of
a base cell are referred to as base edges. Each vertex $x=(i\times
D_0+j\times D_1+k \times D_2)/N$ is represented by $(i,j,k)$ where
$i,j,k\geq 0$, $i+j+k=N$. We call it the {\sl barycentric
coordinates} of the vertex.

By barycentric coordinates, $D_0$ is represented by $(N,0,0)$; $D_1$
by $(0,N,0)$ and $D_2$ by $(0,0,N)$. Boundary vertices along $D_0$
and $D_1$ are the ones in the form $(i,j,0)$ with $i,j>0, i+j=N$.
Other boundary vertices are defined similarly. For any
interior point represented by $(i,j,k)$ we have $i,j,k>0, i+j+k=N$.
Let $V=\{(i,j,k): i,j,k\geq 0, i+j+k=N\}$.

Base cells in the triangulation are oriented in the clockwise order
of their vertices and base edges are oriented according to the
clockwise order of their base cells. See Figure~\ref{orientation} in
Appendix A.1.


A coloring $\phi: V\rightarrow \{0,1,2\}$ is a Sperner
coloring if and only if for any vertex $x =(x_0,x_1,x_2)$,
$\phi(x_0,x_1,x_2)=j\in\{0,1,2\}$ implies $x_j>0$. Sperner Lemma states
that a triangulated triangle with a valid Sperner color has a base cell such
that its three vertices have different colors.

Given a Sperner coloring $\phi:V\rightarrow \{0,1,2\}$ of all
vertices in $V$, let $sign(\delta,\phi)$ and $sign(e,\delta,\phi)$
denote the sign of a base cell $\delta$ and the sign of a base edge
$e$ in $\delta$ respectively. The sign of $e=(u,v)$ is 1 (or -1) if
the colors of its two vertices are $0$ and $1$ and the orientation
of $e$ in $\delta$ is from color $0$ vertex to color $1$ vertex (or
from $1$ to $0$). We denote the sign of a base edge by
$sign(e,\phi)$ if there is no ambiguity on its orientation. In all
other cases, $sign(e,\phi)=0$. The sign of a base cell $\delta$ is
defined to be the sum of the signs of its three base edges.
Therefore,
$sign(\delta,\phi)=sign(e_1,\delta,\phi)+sign(e_2,\delta,\phi)+sign(e_3,\delta,\phi)$.

We may verify the following by a simple case analysis.
\begin{prop}\label{sign}
For any base cell, its sign is $1$ (or $-1$) if and only if its
three vertices are colored with $0,1,2$ in the clockwise
(counterclockwise) order. In all other cases, its sign is zero.
\end{prop}

The index of a connected set of base cells $\Delta$, with respect to
the color $\phi$ is defined as:
$$index(\Delta, \phi)=\sum\{sign(\delta,\phi):\delta\ a\ base\ triangle\in\Delta\}$$
\begin{lemm}~\label{2DSperner}
\cite{Todd} For a triangulated triangle $\Delta$ with colors $\phi:
V\rightarrow \{0,1,2\}$, there are at least $|index(\Delta,\phi)|$
fully colored base cells. The index can be calculated by summing the
signs on its boundary base edges.
\end{lemm}
See Appendix A.2 for the proof. The result also holds for general polygons in 2D.






To generalize the
same result to a higher dimensional polyhedron $P$, we consider a
simpler version of index that is defined mod $2$ and is used in~\cite{CD2006}. For a
$d$-dimensional simplex with vertices assigned $d+1$ different
colors $\{0,1,\cdots,d\}$, we define its index as $1$. Otherwise, it
is defined to be zero. Let $V(P)$ be the vertices of its
triangulation. With respect to a color
$\phi:V(P)\rightarrow\{0,1,\cdots,d\}$, its index is defined as
$index(P,\phi)=\sum_{\delta\in P} index(\delta, \phi)$, where
$\delta$'s are $d$-dimensional simplices in the triangulation of $P$
into simplices.

Denote by $\partial P$ the boundaries of $P$. Note that the
triangulation of $P$ induces a triangulation of $\partial P$ into
$(d-1)$-dimensional simplices. We define $index_{d-1}(\partial P,
\phi)=\sum_{\delta_{ d-1}\in \partial
P}index_{d-1}(\delta_{d-1},\phi)$.

We need the following discrete version~\cite{CD2005} of standard
results on the index defined here.
\begin{prop}~\label{SpernerDegreeTheorem}
$index(P,\phi)\equiv index_{d-1}(\partial P, \phi) \mod 2$.
\end{prop}
See Appendix A.3 for the proof.

\subsection{Kuhn's Triangulation}\label{kuhntri}

In this section, we briefly introduce Kuhn's
triangulation~\cite{Kuhn} for a simplex. Kuhn's triangulation has
the advantage of being a balanced triangulation and it helps us
derive a much improved algorithm for the envy-free cake cutting
problem.

Let us start by explaining the Kuhn's triangulation of a unit cube
in $d$ dimensions. Let $v_0=(0,0,\cdots,0)_{1\times d}$ be one of
the corners of the cube. The diagonal vertex to it would be
$v_{d+1}=(1,1,\cdots,1)_{1\times d}$. Suppose $e_i$ is a
$d$-dimensional unit vector such that $e_{ii}=1$ and $e_{ij} = 0$
for all $i \neq j$. Kuhn's method partitions the cube into $d!$
simplices. Let $\pi:=(\pi(1),\pi(2),\cdots,\pi(d))$ be any
permutation of the integers $0,1,\cdots,d-1$. Each permutation $\pi$
corresponds to one small simplex $\Delta^d_{\pi}$ whose vertices are
given by $v^i_{\pi}=v^{i-1}_{\pi}+{e}_{\pi_{(i)}}$  and
$v^0_{\pi}=v_0$.

These simplices all have disjoint interiors and their union  is the
$d$-cube. It is not difficult to verify this, since any vertex
$x=(x_0,x_1,\cdots,x_{d-1})$  is an interior point of
$\Delta^d_{\pi}$ if and only if
$1>x_{\pi(1)}>x_{\pi(2)}>\cdots>x_{\pi{(d)}}>0$. Appendix A.4
illustrates Kuhn's triangulation on a 3-cube.

Now, we are ready to explain the Kuhn's triangulation of a simplex.
Let $N$ be an integer bigger than $1$.  Take a unit $d$-cube and use
parallel cuts of equal distance to partition it into $N^d$ smaller
$d$-cubes of side length $1\over N$. Then, partition each small cube
into $d!$ simplices using the above method.  Now, observe that the
unit cube can also be partitioned into $d!$ big simplices first and
each big simplex contains $N^d$ smaller simplices or base cells. The
proof of the consistency of the two processes can be found in
Appendix A.5.

Based on the equivalency of the two partitioning processes, we
choose one of the big simplices, for example the one corresponding
to $\pi=(0,1,2,\cdots, d-1)$ and the smaller simplices that are
contained in that will define its triangulation. A vertex $X$ in
this big simplices can be represented by barycentric coordinates
$X=(x_0,x_1,\cdots,x_d)$ by a transformation as illustrated in
Appendix A.6. More details of the transformation can be found in
page 42 of ~\cite{Scarf1977}.

We have the following property of the triangulation:

\begin{lemm}\label{kuhncell}
For given $X=(x_0,x_1,\cdots,x_d)$ and $Y=(y_0,y_1,\cdots,y_d)$,
define $\delta_{X-Y}=\max_{\forall i\in
\{0,1,\cdots,d\}}\{|x_i-y_i|\}$. If  $X$ $Y$ are in the same base cell in Kuhn's
triangulation, then $\delta_{X-Y}=1$.
\end{lemm}

See Appendix A.7 for the proof.

%
%
%
%
%


\section{Finding a Sperner Simplex under Oracle Function Model}

In the oracle function model, the function value at a point (or
color of the point) is given only when it is queried and it remains the same when further queries are
performed on the same point.

We prove that the oracle complexity of finding a Sperner base
simplex under oracle function model in $d$ dimensions is of
$\theta({N^{d-1}})$. Such a matching bound  was known only for
finding a Sperner's fully colored base cell in a two dimensional
$N\times N$ grid~\cite{CS}\cite{FISV}. Our extensions  into higher
dimensions make use of the methodologies originated in the zero
point computation on hypercubes.

The matching bound derived in this section will be essential for solving the envy-free
cake-cutting problem. It could also be of independent
interest for other equilibrium problems that are based on
fixed point computation.

To derive the upper bound, we define the concept of balanced
triangulations:

\begin{defi}
A simplex $P$ is triangulated into balanced simplices of granularity
$g=\frac{1}{N}$ if
\begin{enumerate}
\item $P$ is fully contained in the unit cube
$[0,1]^d$;
\item every parallel plane along the coordinates
$x_i=g\times j$ ($i=1,2,\cdots, d$, $j=0,1,\cdots, N$) cuts through
$P$ along the facets of the base cells of the triangulation, i.e.,
the parallel plane will not cut into the base cells;
\item the number of the $d$-dimensional
simplices of the triangulation within any cube of side length $g$ is
constant.
\end{enumerate}
\end{defi}
By construction, Kuhn's triangulation is a balanced triangulation.

\begin{lemm}\label{simplexupperbound}
For any balanced triangulation, there is an algorithm that finds a
Sperner base cell in time $O(N^{d-1})$ if vertices are colored by a
Sperner coloring.
\end{lemm}
\begin{proof}
We fit the balanced triangulated simplex $P$ into the unit cube
$[0,1]^d$ which guaranteed by condition 1 of Definition 1. By
condition 2, we use parallel plane along the coordinates to cut the
cube.

For the correctness, we note that, which sub-hypercube to look into
will be determined by applying
Proposition~\ref{SpernerDegreeTheorem} on the triangulated simplex.
As the boundary conditions give an odd index for the initial simplex
because of the valid coloring, each time one of the two parts of the
cut simplex will be odd. The procedure can proceed until the last
base cube. Then, we can simply examine up to $C$ remaining simplices
contained in this base cube, where $C$ is a constant (note that
because of $d$ is a constant, the function: $f(d)$ is a constant
function).

For complexity, our algorithm will be doing a binary cut along the
$d$ coordinate one after another. It takes $d$ such cuts to reduce a
hypercube into half of its size(in length). Therefore, in
$d\log_2{(\frac{1}{g})}$ cuts, we reduce the unit hypercube into a
base cube of side length $g$. We upper bound the time complexity
with the total time necessary for the hypercube (the actual number
of operations will be less on the simplex and its triangulation). As
the size reduces geometrically, the total number of operations is
dominated by the number of operations we do at the first $d$ cuts.
For each cut, we need to apply
Proposition~\ref{SpernerDegreeTheorem} to calculate the index of the
boundary simplices of $(d-1)$-dimension, which requires a
computational time $O((\frac{1}{g})^{d-1})$. The computational upper
bound follows.
\end{proof}


We also derive a similar lower bound:


\begin{lemm}~\label{simplexlowerbound}
For any algorithm that finds a Sperner base cell for any
triangulation of a $d$-dimensional simplex with a Sperner coloring,
there exists some input triangulation such that the algorithm takes
time $\Omega(N^{d-1})$.
\end{lemm}

The proof of the above lemma is build on a deep result of Chen and
Deng \cite{CD2005}. It is explained in details in Appendix A.8.

Lemma \ref{simplexupperbound} and Lemma \ref{simplexlowerbound}
result in a matching bound as follows.

\begin{theo}(matching bound)
Given a balanced triangulation where all vertices are colored by
$\{0,1,\cdots, d\}$ by a Sperner coloring,
 a Sperner base cell can be found
in time $\theta(N^{d-1})$.
\end{theo}

\section{Envy-Free Cake Cutting and Sperner Lemma}

As far as we know, the first proof of the existence of envy-free
cake cutting solutions using Sperner's lemma is by
Simmons~\cite{simmons}. Su~\cite{Su} uses a similar argument to
develop a computational procedure to derive an approximate envy-free
cake cutting solution, by a labeling process on barycentric
subdivisions of a simplex (See figure \ref{barysub} as in Appendix
A.9 as well as~\cite{armstrong}\cite{Su}). However, Su's method
creates simplices with large aspect-ratios that make the process
converge rather slowly. Instead, we use Kuhn's triangulation.
%

\subsection{Utility functions and envy-free solutions}
Consider a set $I=\{0,1,\cdots,d\}$ of $d+1$ players. Each player
$i\in I$ has a utility function $u_i$ defined on the Borel space of
the line segment $L=[0,1]$. Our utility functions are required to
satisfy the following two conditions:

\begin{itemize}
  \item Nonnegativity condition: $u_i(\emptyset)=0$ and $u_i(\neq\emptyset)>0$.
  \item Lipschitz condition: For any interval $[x,y]\subseteq L$, $u_i([x,y])\leq K\times |y-x|$.
\end{itemize}

We use $d$ cuts to partition $L$ into a set $S$ of $d+1$ disjoint
segments of lengths $l_0,l_1,\cdots,l_{d}$ such that
$\sum_{i=0}^{d}l_i=1$. Using barycentric
coordinates, we restrict our discussion to integer vectors
$(x_0,x_1,\cdots,x_{d})$ such that
$l_0=x_0/N, l_1=x_1/N, \cdots, l_{d}=x_{d}/N$, with $\sum_{i=0}^d
x_i=N$. All possible partitions form a $d$-dimensional simplex
$\Delta^d$ with $d+1$ vertices. The $i$-th vertex of $\Delta^d$
is represented as $Ne_i$, where $i=\{0,1,\cdots,d\}$ and $e_i$ is
the unit vector whose $i$-th coordinate is 1.


By the nonnegativity condition of utility functions, every
player will strictly prefer the nonzero segments to the zero
segments. Hence, we have the following boundary preference
condition.
\begin{property}
Boundary Preference Property: consider any boundary vertex $X$ that
belongs to a boundary which is incident to the $i$-th corner but not
the $j$-th corner. In the cake cutting defined by $X$, every player
strictly prefers the $i$-th segment to the $j$-th.
\end{property}

The above property actually ensures a Sperner coloring. We can now
give a sketch of the envy-free cake
cutting problem using this connection. The argument has two stages:
labeling and coloring.

For the simplicity of exposition, consider the problem for three
players.  For the case of 3 players, the closed set of all possible cuts
is a triangle. As in the previous section we place $N-1$ equally
spaced line segments parallel to each of the three edges of the
triangle and divide it into $N^2$ equal-size base triangles. Let
$V$ be the set of vertices of all base triangles, i.e.,
$V=\{(x_0,x_1,x_2):x_0,x_1,x_2\geq 0, x_0+x_1+x_2=N\}$.

Next, we partition $V$ into three control subsets $V_0,V_1,V_2$.
Starting by assigning $(N,0,0)$ to $V_0$, $(N-1,1,0)$ to $V_1$, and
$(N-2,2,0)$ to $V_2$. The rest of $V$ is partitioned in such a way
that the three vertices of each base triangle belong to different
subsets $V_t$'s. This can be done by defining $V_t=\{(x_0,x_1,x_2):
x_1-x_2 = t (\mbox{mod~}3) \mbox{ ~for~} x_0,x_1,x_2\geq 0,
x_0+x_1+x_2=N\}$.

Now, we should color $V$. For any vertex $(x_0,x_1,x_2)\in V_t$, we
let player $t$ choose, among three segments,
$[0,l_0],[l_0,l_0+l_1],[l_0+l_1,1]$, of $I$, one that maximizes his
utility. For simplicity of presentation, we should assume a
non-degenerate condition that the choice is unique. The general case
can be handled with a careful tie-breaking rule. If the optimal
segment is the one of length $l_s$, $0\leq s\leq 2$, we assign color
$s$ to the vertex $(x_0,x_1,x_2)$. We claim that the above coloring
is a valid Sperner coloring. This can be easily checked by the
assumption of utility functions. Since $u_i(\emptyset)=0$, the three
vertices $(N,0,0),(0,N,0),(0,0,N)$ of the large triangle must be
colored by $0,1,2$ respectively and the vertices on the edge
$(N,0,0)\rightarrow (0,N,0)$, $(0,N,0)\rightarrow (0,0,N)$, and
$(0,0,N)\rightarrow (N,0,0)$ will be colored either by 0 or 1, 1 or
2 and 2 or 0 respectively by our coloring procedure. Hence, it
satisfies the boundary condition of Sperner lemma, and the coloring
is valid. See Figure \ref{spernerapproach} in Appendix A.10 for an
example.

Since the three vertices of each base triangle belong to three
different subsets, or we say three different players, if we find a
fully colored base triangle, then on the three vertices of this
triangle, different players prefer different segments. By Sperner
Lemma, there exists at least one fully colored base triangle. By
refining the triangulations, the fully colored base triangles
become smaller and smaller, and a subsequence of the base triangles
will converges to a fixed point. Such a fixed point is an envy free
solution for the cake cutting problem. Therefore, there always exist
an envy-free solution for 3 players case.

The case of $d+1$ players for $d > 2$ can be handled in a similar
way. In that case, the closed set of all possible partitions of the
cake forms a $d$-dimensional simplex.

In Kuhn's triangulation the labeling can be done by using
barycentric coordinates. $X=(x_0,x_1,\cdots,x_d)$. For any vertex
$X$ of the base simplex, let $W(X)=\sum_{i=0}^{d}ix_i$. We assign
vertex $X$ to subset $V_t$ if $W(X)\equiv t (\text{mod }d+1)$. This
will partition the vertices of the base simplices into $d+1$ control
subsets $V_0,V_1,\cdots, V_d$. This is a suitable labeling because
$$W(v^i_{\pi})=W(v^{i-1}_{\pi})+((\pi{(i)}+1)-\pi{(i)})=W(v^{i-1}_{\pi})+1.$$

This also proves the existence of a $d$-cut solution.

\begin{theo}~\cite{Stromquist1980,Su}
There is an envy-free cake cutting solution for $d+1$
players that uses only $d$ cuts.
\end{theo}

It is not hard to see that a fully colored base simplex represents an approximate envy-free cake cutting solution.
In fact,  by using the Lipschitz condition defined above, one can
find a cake cutting solution with maximum envy $\epsilon$ through a triangulation in which the sizes
of all base simplices is bounded by $ \epsilon / K$. Motivated by this observation, we define the discrete
cake cutting problem and derive its computation and oracle complexity.


\section{Complexity of Discrete Cake Cutting}

To formalize the analysis for the cake cutting problem, we introduce
a discrete version of the envy-free allocation of a cake among $d+1$
people using $d$ cuts (the d-cut problem for short). We use the
barycentric coordinates $(x_0,x_1,\cdots,x_{d})$ as in the previous section
restricting $x$ to integer vectors
satisfying $\sum_{i=0}^dx_i=N$. We define two d-cuts $x$ and $y$ to
be adjacent to each other, if $\forall i\in\{1,2,\cdots,d\}:
|x_i-y_i|\leq 1$. We call them affine adjacent to each other if they
are adjacent to each other and $|x_0-y_0|\leq 1$.

A discrete cake cut is defined to be a set $\{x^{(0)}, x^{(1)},
x^{(2)},\cdots, x^{(d)} \}$ of $d+1$ $d$-cuts such that for each
pair of $j$ and $k$, the two $d$-cuts $x^{(j)}$ and $x^{(k)}$ are
adjacent. We call it an affine discrete cake-cut if we further
require that $\forall j,k\in\{0,1,\cdots, d\}$: $x^{(j)}$ and
$x^{(k)}$ are affine adjacent.

\begin{defi}
Discrete ENVY-FREE CAKE CUT: A discrete cake cut  is an envy-free
solution if there is a permutation $\pi$ of $\{0,1,\cdots, d\}$ such
that player i prefers the $\pi(i)$-th segment for some d-cut
$x^{(j)}$ in the set. We denote it by $P_i(x^{(j)})=\pi(i)$.
\end{defi}

Note that, our definition is inspired and in line of the definition
of discrete BROUWER fixed point~\cite{PPAD,CD2005,CD2006}.

 \begin{defi}
2D BROUWER: The input is a 2D grid of size $G=N\times N$ ($N=2^n$),
together with a function $f: G\rightarrow \{0,1,2\}$ such that a
boundary condition is satisfied: $\forall y\geq 0: f(0,y)=1$,
$\forall x>0: f(x,0)=2$, and $\forall x,y>0: f(x,N)=f(N,y)=0$. The
required output is a unit square $US=\{(x,y),(x,y+1), (x+1,y),
(x+1,y+1)\}$ such that $f(US)=\{0,1,2\}$.
\end{defi}


Using Kuhn's triangulation, labeling each node can be done in
polynomial time, and so is coloring if the utility functions are
given by a polynomial time algorithm. Therefore, each base cell can
be constructed and their colors verified in polynomial time. In
addition, vertices of each base cell
 in the Kuhn's triangulation are adjacent to each other. Therefore,
the problem reduces to one of finding a fully colored Sperner cell,
which can be done in PPAD. Therefore, we have the following:

\begin{coro}~\label{Inppad}
Finding a discrete d-cut set for Envy-Free Cake Cutting problem for
$d+1$ people is in PPAD.
\end{coro}

On the other hand, we apply a reduction from the 2D BROUWER problem
to prove it PPAD-hard~\cite{2DSperner}; and hence:
\begin{theo}\label{ppad}
Finding an approximate solution for Envy-Free Cake Cutting with $d$
cuts for $d+1$ people is PPAD-Complete.
\end{theo}


\begin{proof}
Given an input function of 2D BROUWER on grid $f: N\times
N\rightarrow\{0,1,2\}$, we embed it into a Kuhn's triangle defined
by three vertices: $<(0,0), (2N,0), (2N,2N)>$ by the mapping:
$M(x,y)=(2N-x,y)$. We define the preference functions (for all
$i=0,1,2$): $P_i(2N-x,y)=f(x,y)$ for $0\leq x,y\leq N$;
$P_i(x,0)=2$, for $0\leq x\leq N$; $P_i(2N, y)=0$ for $N<y\leq 2N$;
$P(x,y)=0$ for all other cases. We name $(0,0)$ the vertex $X_2$,
$(2N,0)$ $X_1$, and $(2N,2N)$ $X_0$. Therefore, the Kuhn's triangle
and the preference functions form a discrete ENVY-FREE CAKE CUT
problem. The boundary condition for the SPERNER is now satisfied and
there is a Sperner colored triangle, which is at the same time a
ENVY-FREE CAKE CUT by our choices of the preference functions.
Therefore ENVY-FREE CAKE CUT does have a solution. Once we find one,
it must be in the region bounded by $N\leq x\leq 2N, 0\leq y\leq N$.
The inverse mapping of $M$ will give us the required BROUWER's
solution.
\end{proof}



Therefore, the envy-free cake-cutting has the same time complexity
as the Sperner Simplex computation if the utility functions are
given by a polynomial time algorithm.

Similar, under the oracle model for utility functions, we should
show the same also holds.

\begin{theo}\label{matchingbound}
Solving the ENVY-FREE CAKE CUT problem of $d+1$ people for the
oracle functions requires time complexity
$\theta(({K\over\epsilon})^{d-1})$.
\end{theo}

\begin{proof}
By Lemma 2, the Kuhn's triangulation in the last section allows us
to find a Sperner's simplex in time $O(({K\over\epsilon})^{d-1})$.
Therefore, the solution corresponds to a discrete ENVY-FREE CAKE CUT
solution.

To prove the lower bound, we apply the same reduction for the
Sperner's problem. Given an input instance of the zero point problem
with direction preserving functions, we introduce the same structure
as before. The only extra definition we need to introduce is for
each vertex $x$ of the grid, we define $\forall i: P_i(x)=j$ if
$f(x)=e_j$ and $P_i(x)=0$ otherwise. Clearly, if a unit square
contains all the preferences, there must be some $i$ such that
$P_i(x)=0$. By direction preserving property, $f(x)$ cannot be
$-e_j$ for any $j$. Therefore $f(x)=0$ and a zero point is found.

Our lower bound of $\Omega(({K\over\epsilon})^{d-1})$ follows.
\end{proof}

\section{A Fully PTAS for 3 Players with Monotone Utility Functions}
For general oracle utility functions, the above results imply a
$\theta({K\over \epsilon})$ matching bound when the number of players is three.
Further improvement can only be possible when we have further
restrictions on the utility functions. In this section, we assume the
utility functions in addition to satisfying the nonnegativity
condition and the Lipschitz condition are also monotone.

Even in this case,  the celebrated Stromquist's moving knife
\cite{Stromquist1980} for envy-free cake cutting of three players
can be shown to have an exponential lower bound (See Appendix A.11).
The main result here is to give an algorithm with
 a running time polynomial in the number of bits of $K$ and
$1\over \epsilon$.

\begin{theo}~\label{polynomialtime}
When the utility functions satisfy the above three conditions, an
$\epsilon$ envy-free solution can be found in time $O(\log^2{K\over
\epsilon})$ when the number of players is three.
\end{theo}

The improvement has been made possible by an efficient way to
compute the index of a triangulated polygon, i.e., counting the sum
of the signs of base intervals on the boundary. The main idea is an
observation that at some appropriate cuts, the colors along the cut
are monotone. Therefore, we can find the boundary of the three
different colors on this cut to calculate the sum quickly.


At any time in the algorithm, we maintain a subset of $V$, with a
non-zero index value, $V(i_1,i_2,k_1,k_2)=\{(i,j,k):i,j,k\geq 0,
i+j+k=N, i_1\leq i\leq i_2, k_1\leq k\leq k_2\}$, delimited by
$i=i_1, i=i_2$ and $k=k_1,k=k_2$.

\begin{algo}
\begin{enumerate}
\item If $i_2-i_1=1$ $\&$ $k_2-k_1=1$, find a fully colored base triangle
of $V(i_1,i_1+1,k_1,k_1+1)$, terminate.
\item Choose $\max \{i_2-i_1,k_2-k_1\}$ (and assume it is $i_2-i_1$ w.l.o.g.).
\item Let $i_3=\lfloor (i_1+i_2)/2\rfloor$.
\item Calculate $index (V(i_1,i_3,k_1,k_2))$ and $index (V(i_3,i_2,k_1,k_2))$.
\item Recurse on one of the two sub-polygons with a non-zero index.
\end{enumerate}
\end{algo}

\noindent By definition, $index(V(i_1,i_2,k_1,k_2))= index
(V(i_1,i_3,k_1,k_2))+index (V(i_3,i_2,k_1,k_2)).$ At least one of
the $index (V(i_1,i_3,k_1,k_2))$ and $index (V(i_3,i_2,k_1,k_2))$ is
non-zero as $index(V(i_1,i_2,k_1,k_2))$ is non-zero. The algorithm
always keep a polygon of non-zero index as the initial polygon has a
non-zero index. At the base case $i_2=i_1+1$ and $k_2=k_1+1$,
$V(i_1,i_1+1,k_1,k_1+1)$ is either already a base triangle (with
non-zero index), or a diamond shape consisting of two base
triangles, one of which must be of index non-zero. The correctness
follows as the only possibility for its index being non-zero is when
it is colored with all three colors.

For complexity analysis, we first characterize the boundary
conditions:
\begin{property}~\label{boundary}
There are up to three types of boundaries for $V(i_1,i_2,k_1,k_2)$
\begin{enumerate}
\item $B_{i=c}=\{(i,j,k)\in V: i ~\mbox{constant}, k_1\leq k\leq k_2, i+j+k=N, i,j,k\geq 0\}$
\item $B_{k=c}=\{(i,j,k)\in V: k ~\mbox{constant}, i_1\leq i\leq i_2, i+j+k=N, i,j,k\geq 0\}$
\item $B_{j=0}=\{(i,0,k)\in V: i+k=N, i,k\geq 0\}$
\end{enumerate}
\end{property}

Second, we establish the monotonicity property. Note that for the third
type of boundaries listed above, the monotonicity property does not hold
for $B_{j=c}$ in general but only for $B_{j=0}$.

\begin{property}~\label{monotone}
The colors of $V_0\cap B_{i=c}$ are monotone in $k$, and so are
$V_1\cap B_{i=c}$ and $V_2\cap B_{i=c}$. The same hold for $V_t\cap
B_{k=c}$, as well as $V_t\cap B_{j=0}$, $t=0,1,2$.
\end{property}

\begin{proof}
Observe that,at $V_0\cap B_{i=c}$, the color is determined by
Player~0 by finding the maximum of $u_0([0,{c\over N}])$,
$u_0([{c\over N}, {c+j\over N}])$, and $u_0([{c+j\over N}, 1])$,
where $0\leq j\leq N-c$. The first item is fixed. The second item is
increasing in $j$ since $[{c\over N}, {c+j\over N}]\subseteq
[{c\over N}, {c+j'\over N}]$ for $j\leq j'$ and $u_i$ is assumed to
be a probability distribution. For the same reason, the last item is
decreasing in $j$. The color will be $0$ if $u_0([0,{c\over N}])$ is
the maximum, $1$ if $u_0([{c\over N}, {c+j\over N}])$ is the
maximum, or $2$ otherwise.  In general, as $j$ increases, the color
in $V_0\cap B_{i=c}$ will start with $2$, then $0$, and finally $1$,
assuming non-degeneracy. However, any of those colors may be
missing.

The same analysis holds for other players and for the case when $k$
is fixed. However, it does not hold when $j$ is fixed, except when
$j=0$.
\end{proof}

Property~\ref{monotone} allows us to find the colors of all the
vertices on the boundaries $V_t\cap B_{i=i_1}$,$V_t\cap B_{i=i_2}$,
 $V_t\cap B_{j=0}$, $V_t\cap B_{k=k_1}$ and $V_t\cap B_{k=k_2}$,
$t=0,1,2$, in  time proportional to logarithm of the length of the
sides, by finding all vertices at which the color changes along it.
Once the changing points of the colors are found, using
monotonicity, the total number of positively and negatively signed
base intervals along the boundary can be calculated in constant
time.


The recursive algorithm reduces the size of $V(i_1,i_2,k_1,k_2)$
geometrically. Let $L=\max\{|i_1-i_2|, |k_1-k_2|)$. $L$ halves in
two rounds of the algorithm. It takes $2\log_2 N$ steps to reduce
$L$ to 1.

At each round of the algorithm, we need to find out, for each of
three players, for each side of $V(i_1,i_2,k_1,k_2)$ (up to five in
all), the boundary of the player's color changes (two boundaries).
That takes time $\log_2 L$, bounded by $\log_2 N$, to derive a total
of $3\times 5\times 2\log_2 N$.

Therefore, the time complexity is $O((\log_2 N)^2)$.

\section{Discussion and Conclusion}

It remains open whether the approximate envy-free cake cutting
problem for
 four or more players would allow for a polynomial time approximation
scheme as in the three player case, when we are dealing with
measurable or monotone functions.
%

\appendix

\section*{Appendix}
\begin{figure}[H]
\centering
\includegraphics[scale=0.4]{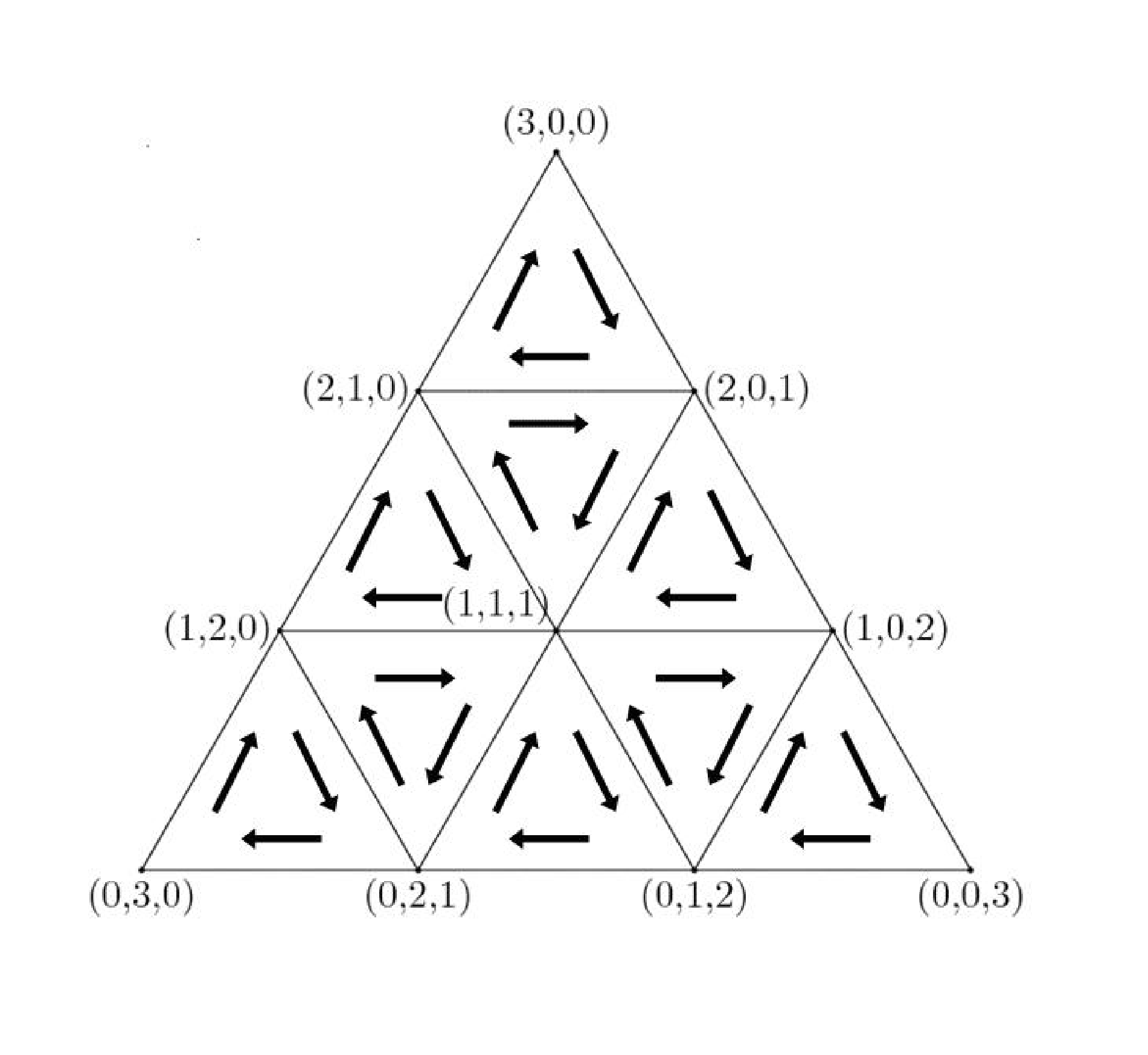}
\caption{Base triangle with edge orientation\label{orientation}}
\end{figure}

\subsection*{A.1 Figure \ref{orientation}: Base triangle with edge orientation}

\subsection*{A.2 Proof of Lemma \ref{2DSperner}}

\begin{proof}
By Proposition \ref{sign} and the definition of $index(\Delta,
\phi)$, the statement that there are at least $index(\Delta, \phi)$
fully colored base triangles is obviously true.

Note that every internal base edge is in two base triangles and has
different signs with respect to the two base triangles. Therefore,
they cancel each other and to derive the following:

$$index(\Delta, \phi)=\sum\{sign(e,\phi):e\ a\ boundary \ base\
edge\}$$

Therefore, we complete the proof.

Note that in the above sum, the orientation of the boundary base
edges are in the clockwise order around $\Delta$.
\end{proof}

\subsection*{A.3 Proof of Proposition \ref{SpernerDegreeTheorem}}

\begin{proof}
 For each $d$-dimensional
simplex $index_d(\delta,\phi)$ is one if and only i
f the set of
colors of its $d+1$ vertices are all distinct and is the same as
$\{0,1,\cdots, d\}$. On the other hand, it has $d+1$ faces of
$d-1$-dimension. Each face of $(d-1)$-dimension is a simplex of
$(d-1)$-dimension. By the definition of the induced index
$index_{d-1}$ only considers colors $\{0,1,\cdots,d-1\}$, a
$(d-1)$-dimension simplex has an index $1$ if and only if the set of
colors of all its vertices is the same as $\{0,1,\cdots,d-1\}$. Now
there is only one vertex left for which we don't know its color. If
it is anything in $\{0,1,\cdots,d-1\}$, we have exactly one more
face having index equal to 1. In that case, summing up the indices
of the faces, we obtain a sum of two which is zero mod 2. If the
last color is $d$, then the sum will be one. Therefore, the claim
holds when $P$ is a simplex with no further triangulation.

In general, $index(P,\phi)=\sum_{\delta\in P} index(\delta,\phi)$.
We can replace $index(\delta,\phi)$ by the sum of indices of the
boundaries of $\delta$. However, each $(d-1)$-dimensional simplex in
the triangulation appears in exact two $d$-dimensional simplices in
the triangulations, unless it is at the boundary of $P$ where it
appear only once. Since we consider the sum mod 2, all the terms
cancel out except those on the boundaries of $P$. The claim follows.
\end{proof}

\subsection*{A.4 An example of Kuhn's triangulation for a 3-cube}

Let $(0,0,0)$ be the base point, according to the different
permutations of 0,1,2, we obtain six tetrahedrons which is a
simplicial partition of the unit cube.

$$\pi=(0,1,2): \Delta= \{(0,0,0),(1,0,0),(1,1,0),(1,1,1)\};$$
$$\pi=(0,2,1): \Delta= \{(0,0,0),(1,0,0),(1,0,1),(1,1,1)\};$$
$$\pi=(1,0,2): \Delta= \{(0,0,0),(0,1,0),(1,1,0),(1,1,1)\};$$
$$\pi=(1,2,0): \Delta= \{(0,0,0),(0,1,0),(0,1,1),(1,1,1)\};$$
$$\pi=(2,0,1): \Delta= \{(0,0,0),(0,0,1),(1,0,1),(1,1,1)\};$$
$$\pi=(2,1,0): \Delta= \{(0,0,0),(0,0,1),(0,1,1),(1,1,1)\}.$$

\begin{figure}[H]
\centering
\includegraphics[scale=0.6]{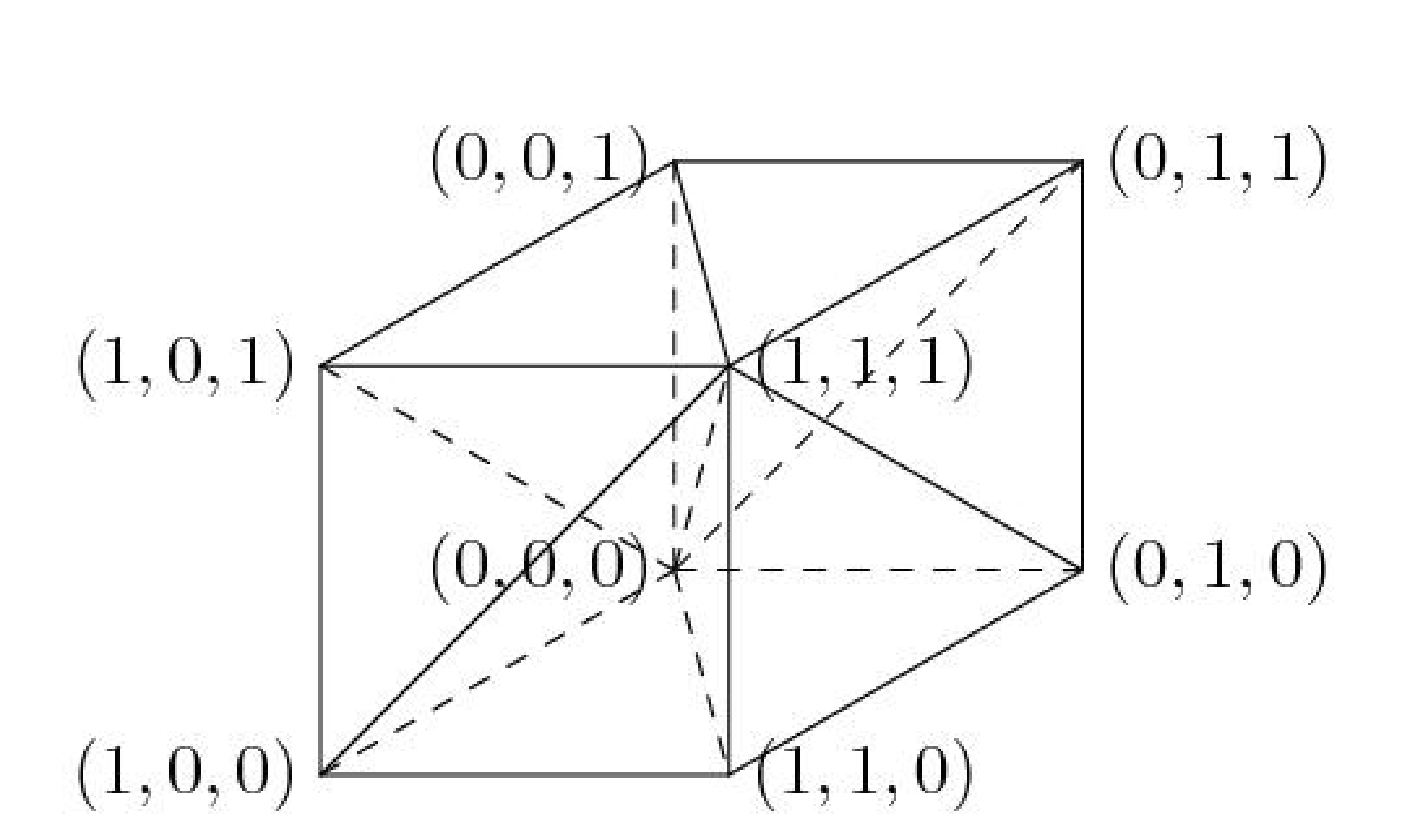}
\caption{An illustration of Kuhn's partition on a unit cube in 3
dimension\label{kuhn}}
\end{figure}

The partition is illustrated in Figure \ref{kuhn}.

\subsection*{A.5 Proof of the consistency of the two processes}
we scale the unit cube into one of side length $N$. We call it the
big cube, and each of the $N^d$ sub-cubes as the small cubes.

It is enough to show that, for a simplex in the refined grid
starting from any grid point $x^*=(x^*_1, x^*_2,\cdots, x^*_d)$ to
 $x^*=(x^*_1+1, x^*_2+1,\cdots, x^*_d+1)$,
defined by a permutation $\gamma$, all its $d+1$ vertices are in the
same big simplex derived by some permutation $\rho$
 with base point
$(0,0,\cdots,0)$ to the point $(N,N,\cdots,N)$.

Let $x^0=x^*$, and $x^i$ is the same as $x^{i-1}$ except in its
coordinate $x^i_{\gamma(i)}$ which is $x^{i-1}_{\gamma(i)}+1$.
Therefore, if $x^*_i>x^*_j$, then $x^*_i\geq x^*_j+1$ and then
$x^{k}_i\geq x^{k}_j$ for all $k=1,2,\cdots, d$. Similarly, if
$x^*_i<x^*_j$, then $x^{k}_i\leq x^{k}_j$ for all $k=1,2,\cdots, d$.
If $x^*_i=x^*_j$, then we have $x^k_i\geq x^k_j$ for all
$k=1,2,\cdots, n$, or $x^k_i\leq x^k_j$ for all $k=1,2,\cdots, n$,
dependent on $\gamma(i)<\gamma(j)$ or $\gamma(i)>\gamma(j)$.

Therefore, there is a permutation $\rho$ such that
$x^k_{\rho(1)}\geq x^k_{\rho(2)}\geq \cdots \geq x^k_{\rho(d)}$ for
all $k=0,1,2,\cdots,d$, which guarantees the base simplex inside one
of the $d!$ large simplices.

\subsection*{A.6 Transformation process}
This can be done by a transformation as follows. Let $X$ be a base
point in barycentric coordinates. Set
$e_i=(e_{i0},e_{i1},\cdots,e_{id})$ where $e_{ij}=0$ for all $i$ and
$j$ except that $e_{ii}=-1$, $e_{i,i+1}=1$ for $i=0,\cdots,d-1$.
Then the vertices of the base simplex according to permutation $\pi$
based on $X$ are given by $v^i_{\pi}=v^{i-1}_{\pi}+{e}_{\pi_{(i)}}$
for $\forall i=1,\cdots,d$ and $v^0_{\pi}=X$. For example, for a
simplex based on vertex $v^0=(0,0,\cdots,0)_d$ corresponding to
permutation $(0,1,2,\cdots,d-1)$. We first set
$v^0=(1,0,\cdots,0)_{d+1}$, then by the above transformation
$v^1=(0,1,\cdots,0)_{d+1}$, $v^2=(0,0,1,\cdots,0)_{d+1}$. Please
refer to page 42 of ~\cite{Scarf1977} for more details.

\subsection*{A.7 Proof of Lemma \ref{kuhncell}}
\begin{proof}
WLOG, assume the base cell corresponds to a permutation $\pi$ and
$Y=X+\sum_{i=l}^k e_{\pi_i}$ for some $0\leq l\leq k\leq d$. Fix
$i$, since $e_{ii}=1$, $e_{i,i+1}=-1$ are the only non-zero
coordinates in $e_i$, let $a=(a_1,a_2,\cdots,a_{d+1})=\sum_{i=l}^k
e_{\pi_i}$, then $\forall i$, $a_i\in \{-1,0,1\}$. Therefore,
$\delta_{X-Y}$ equals to either 0 or 1 and since $X$, $Y$ are two
different vertices, we must have $\delta_{X-Y}=1$.
\end{proof}

\subsection*{A.8 Proof of Lemma \ref{simplexlowerbound}}
\begin{proof}
We establish a reduction from the zero point problem considered by
Chen and Deng~\cite{CD2005} for the direction preserving functions.
to the Sperner problem of dimension $d$ to derive a similar matching
bound within a constant factor for any constant dimension $d$.

We achieve the goal in two steps. Let $N=\frac{1}{g}$. First, we
reduce the problem in~\cite{CD2005} to one of finding a discrete
fixed point over a base hypercube on hypergrids for functions from
the grid points $N^d$ to $\{0,1,2,\cdots, d\}$ satisfying the usual
boundary conditions, where a discrete fixed point over a base
hypercube ($2^d$ points within distance one in $|\cdot|_\infty$
metric) is one with function values on the nodes of the base
hypercube cover all the values from $0$ to $d$.

To do that, we quickly review the fixed point problem for direction
preserving functions discussed in~\cite{CD2005}. We consider
functions which are defined on the grids point $N^d$, which has
values $U_d:=\{0,\pm e_1, \pm e_2,\cdots, \pm e_d\}$, where $e_i$ is
the unit vector with the $i$-th coordinate being $1$. A function $f:
N^d\rightarrow U_d$ is direction preserving if and only if for each
$x,y\in N^d$ with $|x-y|_\infty\leq 1$, $f(x)^T\cdot f(y)\geq 0$.
That is, the function values of nodes within distance $1$ of each
other cannot be of different signs. We define a function
$h:N^d\rightarrow \{0,1,\cdots,d\}$ such that $h(x)=0$ if $f(x)\leq
0$ and $h(x)=i$ if $f(x)=e_i$. If we find
 a discrete fixed point set in $h$ on a base hypercube, then
the base hypercube must contain a node $x$ such that $f(x)=0$ as $f$
is direction preserving. Therefore, finding a zero point in $f$ can
be reduced to the problem of finding a discrete fixed point of the
hypercube form in function values of $h$. There is still an issue
whether such a solution exists. For the zero point problem, in
addition to the requirement of direction preserving, it is assumed
that the function $f$ is bounded, i.e., $f(x)+x\in N^d$.  In fact,
we can further assume a specific boundary condition that either
$f(x)=e_i$ if $x_i=0$ and $\forall j<i: x_j>0$, $f(x)=-e_i$ if $x>>0
$ (all the coordinates of $x$ are positive), $x_i=N$ and $\forall
j<i: x_j<N$. In fact, for any bounded function $f$, we can always
cover it with one more layer on each face of its boundary (and
function values) to achieve that. It is easy to verify the direction
preserving condition still holds for bounded functions satisfying
the property in the original cube. Now it is easy to see there is
only one cube of dimension $(d-1)$ that has an index one on the
boundary. By Proposition \ref{SpernerDegreeTheorem}, there must be a
base simplex inside the hypercube that has index one, i.e., has its
vertices colored differently from $\{0,1,\cdots, d\}$.

\begin{figure}[H]
\centering
\includegraphics[scale=0.5]{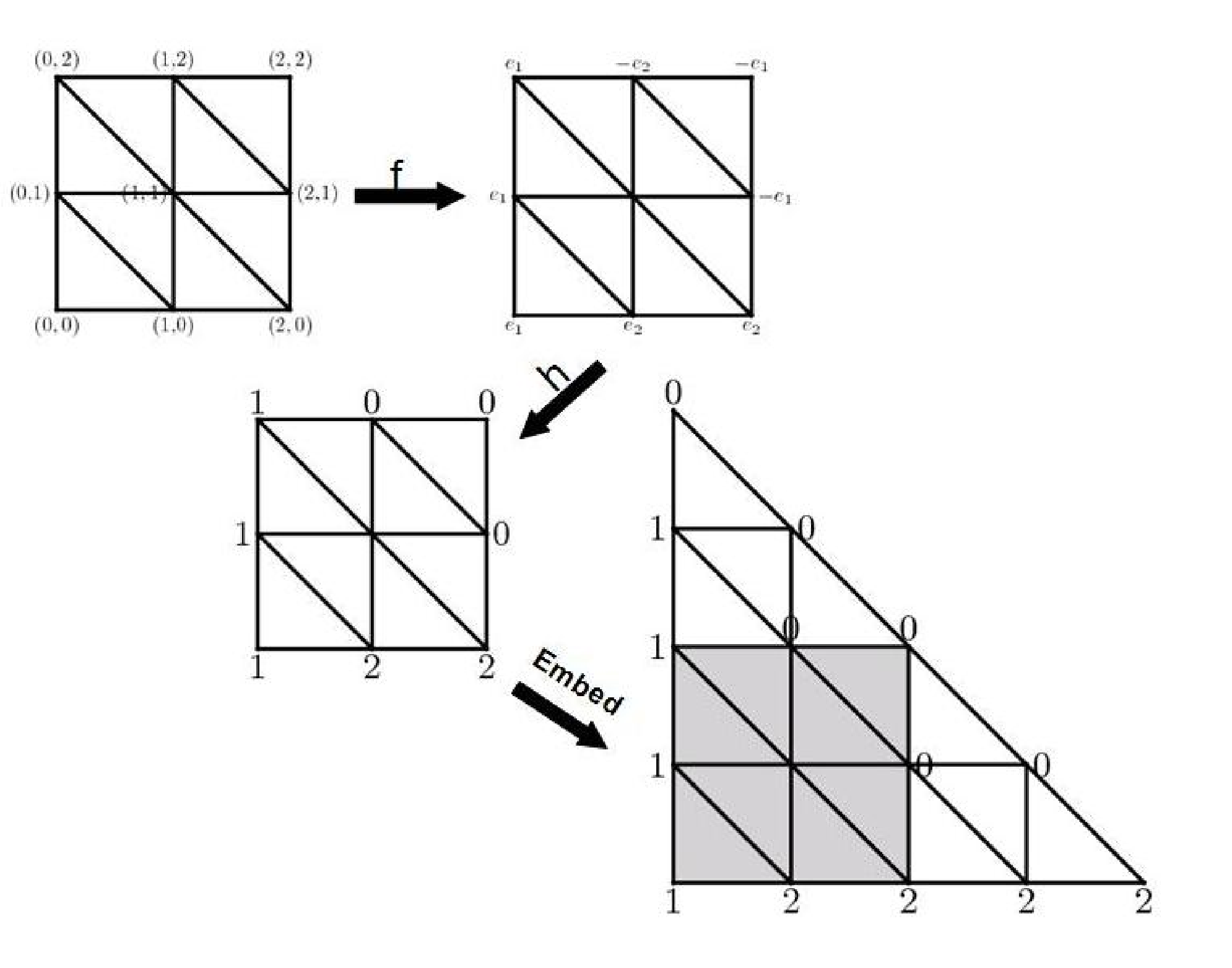}
\caption{Reduction process in Lemma 3\label{lowerboundfig}}
\end{figure}

In the second (and last) step, we  note that there is a size
$(\frac{N}{d})^d$ hypergrid in a $d$ dimensional simplex of length
$N$. The lower bound $\Omega{((\frac{N}{d})^{d-1})}$ follows from a
lower bound of $\Omega{(N^{d-1})}$ for the fixed point problem.
Since $d$ is a constant here, we obtain the lower bound
$\Omega{(N^{d-1})}$. We need to add that, the hypergrid can be
embedded into the simplex such that its coloring can be extended to
a valid coloring of vertices in the triangulated simplex. We align
the colored hypergrid in a way that the origin is placed at the
origin of the simplex, and align the $d$ rays out of the hypergrid
with the rays of the simplex out of its origin. For the overlapping
vertices of the simplex, they are colored with the same colors as in
the hypergrid. The faces of the simplex passing through the origin
will be colored by the same rules as the corresponding faces of the
hypergrid, except the vertices $Ne_1, Ne_2, \cdots, Ne_d$. which are
colored in the following rules: $Ne_i$ is colored with $i+1$ and
$Ne_d$ is colored with $0$. Note that the origin is colored with
$1$. All the extra vertices of the simplex will be colored $0$. This
way, the boundary conditions of the simplex are satisfied. In
addition, no new fully colored base simplex is introduced into the
triangulated simplex. See Figure \ref{lowerboundfig}.

\end{proof}

\subsection*{A.9 Figure \ref{barysub}: Barycentric subdivision on a two dimensional triangle}
\begin{Figure}[H]{Barycentric subdivision on a two dimensional triangle}[barysub]
    \graphfile[33]{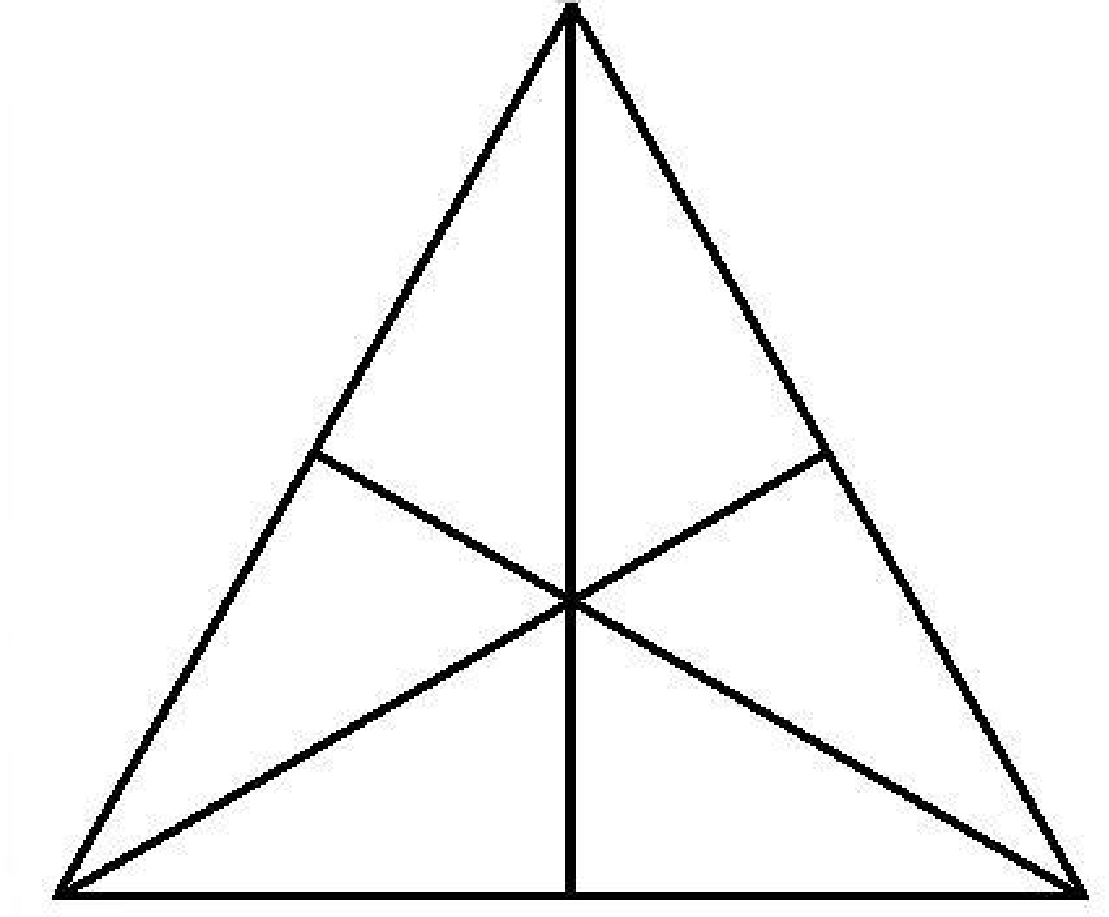}[1st iteration]
    \graphfile[33]{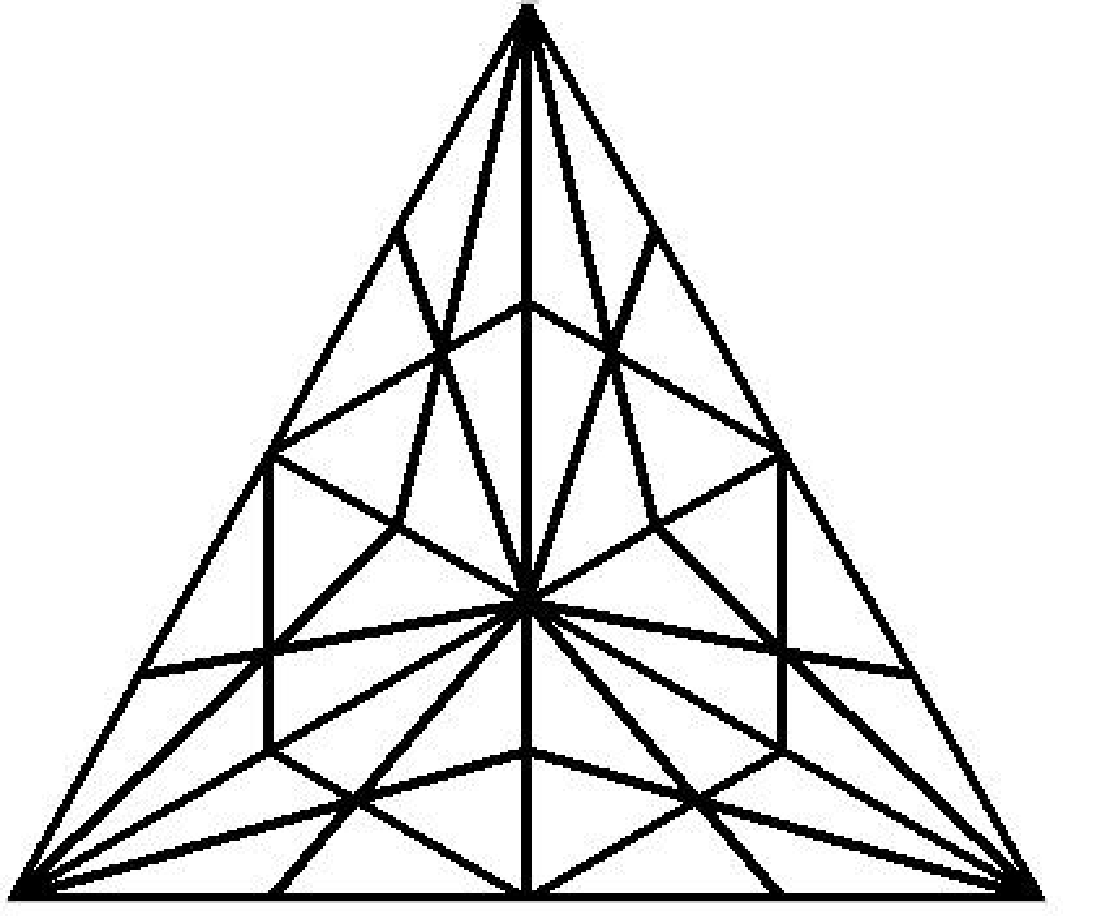}[2nd iteration]
    \graphfile[34]{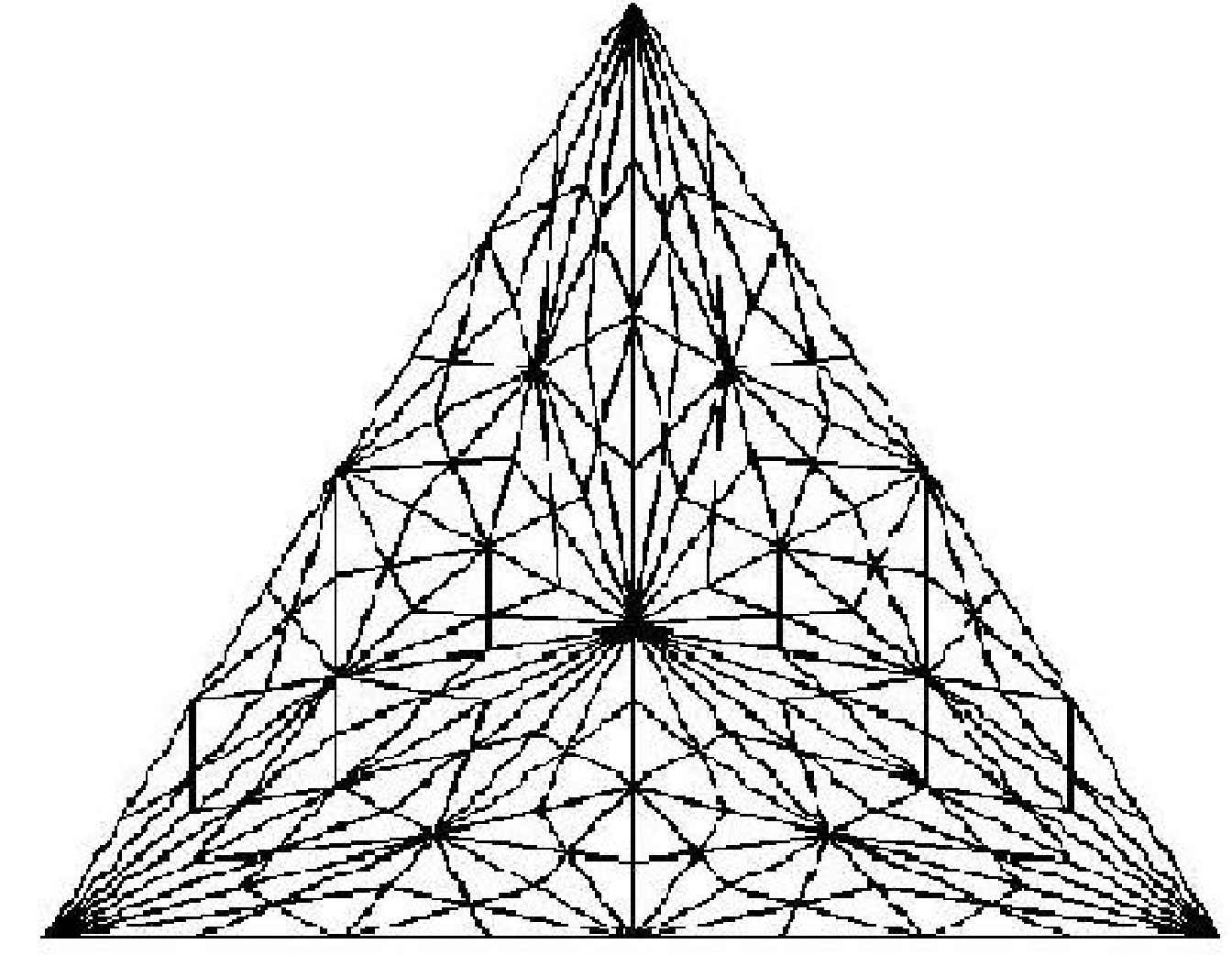}[3rd iteration]
\end{Figure}

\begin{figure}[H]
\centering
\includegraphics[scale=0.5]{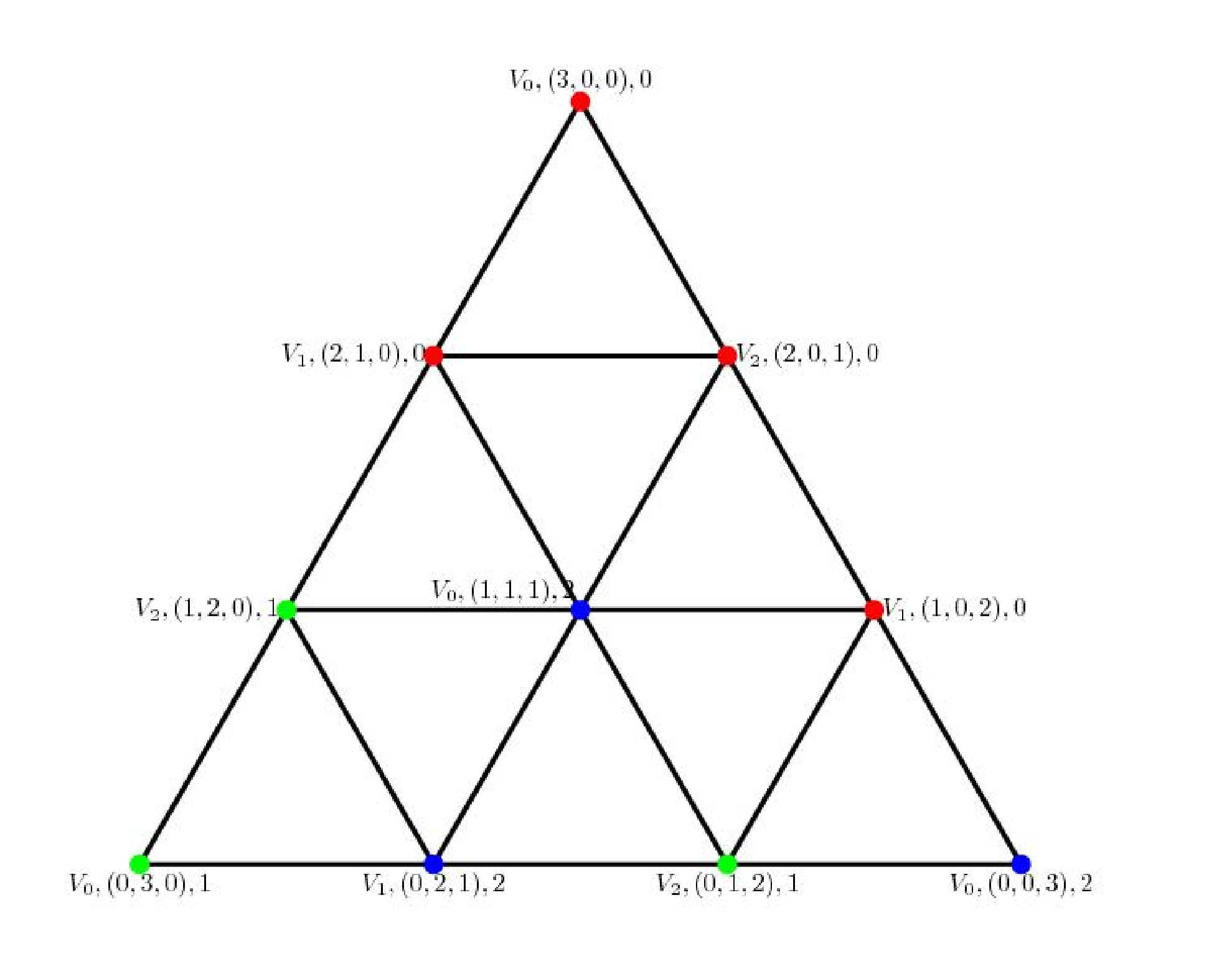}
\caption{Sperner simplex approach for 3 players envy-free
cake-cutting\label{spernerapproach}}
\end{figure}

\subsection*{A.10 Figure \ref{spernerapproach}: An illustration of Sperner simplex approach for 3 players envy-free cake-cutting}

\subsection*{A.11 Proof of an exponential lower bound for Stromquist's solution}
The celebrated Stromquist's solution \cite{Stromquist1980} involves
a referee who moves her sword from left to right. The three players
each has a knife at the point that would cut the right piece to the
sword in half, according to their own valuation. While the referee's
sword moves right, the three knives all move right in parallel but
possibly at different speeds. At all times, each player evaluates
the piece to the left of the sword, and the two pieces that would
result if the middle knife cuts. If any of them sees the left piece
of the sword is the largest, he would shout "cut". Then the sword
cuts. The leftmost piece is assigned to the player who shouted. For
the two players who didn't shout, one whose knife is to the left of
the middle knife receives the middle piece, and one whose knife is
on the right of the middle knife receives the rightmost piece.

we will show that Stromquist's moving knife procedure can not be
turned into a polynomial time algorithm for finding an
$\epsilon$-envy free solution. We will do this by showing that the
particular fixed point found by the Stromquist's procedure can not
be found with polynomial number of queries. We assume a query can
ask about the valuation of a player for an interval $(x,y)$.

Suppose we have three players A, B, and C. The cake is represented
as an interval $[0,1]$. Players A and B have the same utility
function. Their utility functions can be described as
follows~\footnote{We should normalize them for consistency but did
not for simplicity of presentation.}.

\begin{equation}
u_A(x,y) = u_B(x,y) =
\begin{cases}
0 & \mbox{for~} x = 0 , y = 1/10 \\
2 & \mbox{for~}x = 1/10 , y = 3/10 \\
100 & \mbox{for~} x = 3/10, y = 4/10 \\
2 & \mbox{for~} x = 4/10, y = 8/10 \\
100 & \mbox{for~} x = 8/10, ~y = 9/10 \\
0 & \mbox{for~} x = 9/10, ~y = 1 \\
\end{cases}
\end{equation}

The value of both players for any other interval can be computed
assuming that their valuation is uniform across all the above
intervals. For example $u_A(1/20, 3/20) = 0.5$. $u_C$ can be
described in a similar way as follows:

\begin{equation}
u_C(x,y) =
\begin{cases}
100 - \delta & \mbox{for~} x = 0 , y = 1/10 \\
2 & \mbox{for~}x = 1/10 , y = 3/10 \\
98 & \mbox{for~} x = 3/10, y = 4/10 \\
2 & \mbox{for~} x = 4/10, y = 5/10 \\
0 & \mbox{for~} x = 5/10, ~y = 1 \\
\end{cases}
\end{equation}

Assume that the value of $\delta$ is very small. Now, consider the
Stromquist's moving knives. Suppose the referee starts by moving her
knife from $0$ towards $1$. A and B have the same utility function
so their knives are going to be at the same place. Moreover, the
"middle knife" will be always A's or B's. Observe that when the
referee's knife reaches point $1/10$ the middle knife is at point
$4/10$. No one will shout cut yet. Now, observe that since according
to A and B the density of the interval (1/10, 3/10) is twice the
interval (4/10, 5/10), the speed of the knives of A and B will be
exactly the same as the referee's knife. For a similar reason,
according to C the value of the middle piece (from referee's knife
to the knife of A or B) will remain just $\delta$ above the value of
the leftmost piece until the referee's knife goes slightly beyond
$3/10$. At that time C will shout cut and receive the leftmost
piece. Players A and B happily split the rest of the cake equally.

Now, we will slightly perturb the utility function of C. At a point
$1/10 \leq x \leq 3/10$ perturb the utility function of C in the
following way: increase the utility of C for the interval $( x -
\delta, x)$ by $\delta/2$ and decrease his utility for the interval
$(x, x + \delta)$ by the same amount. Call the player with this
perturbed utility function $C_x$.

Now, it is easy to see that if we are running the Stromquist's
method, player $C_x$ will shout cut at the time referee's knife
reaches $x$. It is also not hard to see that any value query asking
for the value of an interval can not distinguish $C$ from $C_x$
unless one end points of the interval for which it is querying the
value is in $(x - \delta, x + \delta)$. Therefore, it is not
possible to distinguish $C_x$ from $C$ with queries of order
polylogarithmic in $1/\delta$.

\end{document}